\documentclass[twocolumn,aps,pre,amsmath,amssymb,longbibliography]{revtex4}
\usepackage{graphicx}
\usepackage{dcolumn}
\usepackage{bm}
\usepackage{multirow}
\usepackage{amssymb}
\usepackage{amsmath}
\usepackage{graphicx}
\usepackage{xcolor}
\usepackage[colorlinks,citecolor=blue,linkcolor=red,urlcolor=blue]{hyperref}
\usepackage{indentfirst}
\usepackage{amsmath}
\usepackage{cases}

\begin{document}
\title{Hybrid scaling properties of the localization transition in a non-Hermitian disordered Aubry-Andr\'{e} model}
\author{Yue-Mei Sun}
\author{Xin-Yu Wang}
\author{Liang-Jun Zhai}

\email{zhailiangjun@jsut.edu.cn}
\affiliation{The school of mathematics and physics, Jiangsu University of Technology, Changzhou 213001, China}

\date{\today}
\begin{abstract}
In this paper, we study the critical behaviors in the non-Hermitian disorder Aubry-Andr\'{e} (DAA) model, and we assume the non-Hermiticity is introduced by nonreciprocal hopping.
We employ the localization length $\xi$, the inverse participation ratio ($\rm IPR$), and the energy gap $\Delta E$ as the characteristic quantities to describe the critical properties of the localization transition.
By performing scaling analysis, the critical exponents of the non-Hermitian Anderson model and the non-Hermitian DAA model are obtained,
and these critical exponents are different from their Hermitian counterparts, indicating that the Hermitian and non-Hermitian Anderson and DAA models belong to different universality classes.
The critical exponents of the non-Hermitian DAA model are remarkably different from both the pure non-Hermitian AA model and the non-Hermitian Anderson model, showing that disorder is an independent relevant direction at the non-Hermitian AA model critical point.
We further propose a hybrid scaling law to describe the critical behavior in the overlapping critical region constituted by the critical regions of the non-Hermitian DAA model and the non-Hermitian Anderson localization.
\end{abstract}
\maketitle

\section{Introduction}

In the process of material preparation and experiment, disordered factors such as impurities and defects are inevitable.
In 1958, Anderson proposed the famous Anderson model to investigate the effect of disorder on phases and phase transitions~\cite{Anderson1958}.
Theoretically, one dimensional (1D) or 2D Anderson model localizes for any infinitesimal disorder amplitude, and in 3D Anderson model the localization transition should emerge at some finite nonzero disorder amplitude~\cite{Jan2023,Alexey2023}.
In addition to disorder, quasiperiodic systems~\cite{Agrawal2020,Goblot2020,Roy2022,Agrawal2022,Zhou2023} where the translational invariance is broken by the incommensurate period can also lead to Anderson localization.
Among many theoretical quasiperiodic models, the Aubry-Andr\'{e} model (AA) is one of the most celebrated examples ~\cite{Aubry1980,Biddle2011,Xu2019,Ting2022,Xuan2022,Xuan2023,Ganeshan2015,Strkalj2021,Purkayastha2018,Sutradhar2019,Khemani2017,HepengYao2019,Yang2017}, partly inspired by its realization in the pseudorandom optical lattice~\cite{Roati2008} and ultracold atoms~\cite{Billy2008}.
A remarkable feature of the AA model is self-duality, which manifests an energy independent localization transition occurring at finite quasi-periodic potential\cite{Biddle070601}.

Although both disorder and quasi-periodicity can lead to localization transitions, theoretical studies have uncovered significant disparities between the two mechanisms, e.g., the localization-extended transition can happen even in the 1D AA model~\cite{Aubry1980}.
Moreover, scaling analysis showed that localization transitions in disorder systems and in quasiperiodic systems belong to two distinct universality classes and present different critical behavior~\cite{Sinha2019,Wei2019,Cestari2011,Xuan2022,Agrawal2020,XLuo2021, Prasad094204,Khemani2017,zhangsx2018,Hamizaki2019,zhai2020}.
For example, the critical exponents of the 1D AA model and the 1D Anderson model are different~\cite{Sinha2019,Wei2019,Cestari2011,Xuan2022}.
For the many-body localization (MBL) in interacting disorder or quasiperiodic systems, exact diagonalization and real space renormalization group studies indicated that MBL transitions can also be categorized into different universality classes with their specific critical exponents which depend on symmetry class and dimensionality~\cite{Khemani2017,zhangsx2018,Hamizaki2019,zhai2020}.
Recently, Bu \textit{et al.} have introduced a disorder AA (DAA) model that seamlessly merges the mechanisms of localization into a unified framework~\cite{Xuan2022,Xuan2023}.
It was shown that disorder and quasiperiodic potentials act as two different relevant directions, and rich critical phenomena in the critical region spanned by the quasiperiodic and disorder potentials were found.
In particular, a remarkable characteristic of this model is the presence of an overlapping critical region constructed from the critical regions of the DAA model and Anderson localization.

In recent years, there has been a surge in the development of non-Hermitian physics, which has found applications across a broad spectrum of condensed matter physics~\cite{Mostafazadeh2001,El-Ganainy2018,Yuto2021,Miri7709, Kunst026808,Okuma033133,Zhang2109431,Kawabata041015,Borgnia056802,Aodong706,Yao2018,Yao136802,Guo2021}.
The interplay between non-Hermiticity and disorder or quasiperiodicity also brings new perspectives to our understanding of localization transition~\cite{zhai2020,Hatano1996,Hatano1998,Cheng2399,Wang024202,Luo0153523,Tzortzakakis014202,ZHANG2023157,Zhang267062,Longhi2019,Longhi20192,Cai2021,
Jiang2019,Liu2021,Liu20201,Chen144208,Wang024514,Longhi224206,Jazaeri2001,PWang2019,zhai2021,Zhai2022,Tang2021,Chaohua123048,Zeng2024,Martinez2018}.
The non-Hermitian extension of the Anderson model by Hatano and Nelson, discovered that a mobility edge can indeed form even in 1D~\cite{Hatano1996,Hatano1998}.
For the non-Hermitian quasiperiodic system, it has been demonstrated that non-Hermiticity can induce reentrant localization, i.e., the localization transition can appear twice as the strength of the quasiperiodicity is increased~\cite{Chaohua123048}.
For the non-Hermitian AA model with nonreciprocal hopping or gain and loss, it has been observed that the localization transition consistently occurs in tandem with both a topological phase transition and a transition from real to complex of energy spectra~\cite{Jiang2019,Longhi2019,Longhi20192}.
More importantly, the introduction of non-Hermiticity can also significantly alter the critical behavior of localization transitions, e.g., the Hermitian and non-Hermitian AA systems belong to different universality classes~\cite{Sinha2019,Zhai2022}.
However, the impact of non-Hermiticity on the critical behaviors of the DAA model still remains unexplored.

In this paper, we investigate the scaling properties of the localization transition in a non-Hermitian DAA model,
and we assume that the non-Hermiticity of this model is introduced by nonreciprocal hopping.
We use the localization length $\xi$, the inverse participation ratio (IPR), and the energy gap $\Delta E$ (i.e. the difference between the second lowest value and the lowest value of the real part of eigenenergy) as characteristic observables to perform our scaling analysis.
The scaling functions of these quantities are established, and the critical exponents for the pure non-Hermitian Anderson model and the non-Hermitian DAA model are determined.
We find that the exponents of the non-Hermitian DAA model are different from those of both the non-Hermitian Anderson model and the non-Hermitian AA model, indicating that the disorder is an independent relevant direction at the non-Hermitian AA critical point.
Our scaling analysis also discovers that the non-Hermitian DAA model and the Hermitian DAA model belong to different universality classes.
Furthermore, an overlapping critical region, constituted by the critical regions of the non-Hermitian DAA model and the non-Hermitian Anderson localization transition, is also found for the non-Hermitian DAA model, and a hybrid scaling law for localization transition in this overlapping critical region is proposed.

The rest of the paper is arranged as follows.
The non-Hermitian DAA model and the characteristic observables are introduced in Sec.~\ref{secmodel}.
In Sec.~\ref{criticalexponent}, we perform our scaling analysis on the pure non-Hermitian Anderson model and the non-Hermitian DAA model, and determine the critical exponents.
Then in Sec.~\ref{Hybridscaling}, by taking disorder and quaisperiodic potentials as scaling variables, the general finite size scaling forms of these three observables are established and verified. Moreover, a hybrid scaling law in the overlapping region is also proposed and numerically verified.
A summary is given in Sec.~\ref{sum}.

\section{\label{secmodel}The non-Hermitian DAA Model and the character observables }
\subsection{The non-Hermitian DAA Model}
The Hamiltonian of the non-Hermitian DAA model reads
\begin{eqnarray}
\label{Eq:model}
H &=& -\sum_{j}^{L}{(J_L c_{j}^\dagger c_{j+1}+J_Rc_{j+1}^\dagger c_{j}})+\Delta \sum_{j}^{L}w_j c_j^\dagger c_j\\ \nonumber
&&+(2J_R+\delta)\sum_{j}^{L}\cos{[2\pi(\gamma j+\phi)]c_j^\dagger c_j},
\end{eqnarray}
in which $c_j^\dagger (c_j)$ is the creation (annihilation) operator of the hard-core boson, $J$ represents the hopping coefficient, $J_L=Je^{-g}$ and $J_R=Je^{g}$ are the asymmetry hopping coefficients, $w_j\in[-1,1]$ gives the quenched disorder configuration, and $\Delta$ measures the disorder strength;
$(2J_R+\delta)$ measures the amplitude of the quasiperiodic potential, $\gamma$ is an irrational number, and $\phi\in[0,1)$ is the phase of the potential.
The periodic boundary condition (PBC) is imposed in the following calculation.
To satisfy PBC, $\gamma$ has to be approximated by a rational number $F_n/F_{n+1}$ where $F_{n+1}=L$ and $F_{n}$ are the Fibonacci numbers~\cite{Jiang2019,Zhai2022}. In the following, we assume $J=1$ as the unit of energy.

For the non-Hermitian AA model, i.e., $\Delta=0$ in Eq.~(\ref{Eq:model}), previous studies have found that it undergoes a localized-extended phase transition at $\delta=0$~\cite{Zhai2022,Jiang2019}.
For the non-Hermitian Anderson model, i.e., $\delta=-2J_R$ in Eq.~(\ref{Eq:model}), its state with the lowest real part of the eigenenergy remains localized at any finite values of $\Delta$, indicating that the phase transition point is always located at $\Delta=0$~\cite{Hatano1996,Hatano1998}.

\begin{figure}[tbp]
\centering
  \includegraphics[width=0.8\linewidth,clip]{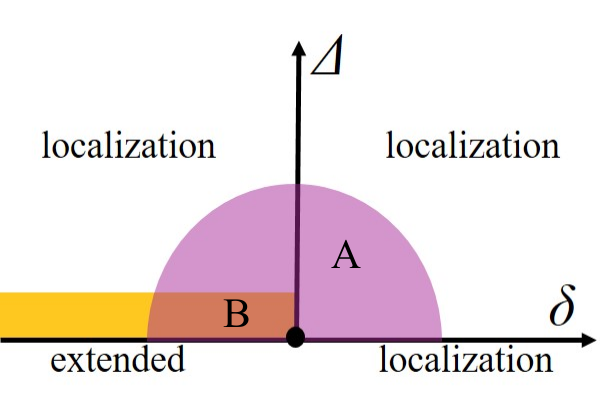}
  \vskip-3mm
  \caption{Sketch of the phase diagram of the non-Hermitian DAA model. The region A (violet region) denotes the critical region of localization transition of the non-Hermitian DAA model.
  The region B (yellow region) denotes the critical region of the Anderson localization transition.
  The intersection of regions A and B represents the overlapping critical region where non-Hermitian DAA and non-Hermitian Anderson localization critical regions coexist. }
  \label{phase}
\end{figure}

For the non-Hermitian DAA model, we have delineated the phase diagram within the $\delta$-$\Delta$ parameter plane under a specified value of $g$, as depicted in Fig.~\ref{phase}.
When $\delta>0$, the non-Hermitian DAA is in the localized phase.
When $\delta<0$ and $\Delta=0$, this model reverts to the non-Hermitian AA model, and all the eigenstates are extended.
For the pure 1D non-Hermitian Anderson model, the Anderson transition of the state with the lowest real part of the eigenenergy occurs at $\Delta=0$ when $L\rightarrow\infty$,
which means infinite disorder will localize the wave function for $\delta<0$.
Around the critical point $(\delta,\Delta)=(0,0)$, the critical region of the non-Hermitian DAA model is spanned by $\Delta$ and $\delta$.
For $\delta<0$ and infinitesimal $\Delta$, there is a critical region of Anderson localization.
As a result, near the critical point $(\delta,\Delta)=(0,0)$ and $\delta<0$, these critical regions inevitably overlap with each other.

\begin{figure}[tbp]
\centering
  \includegraphics[width=1\linewidth,clip]{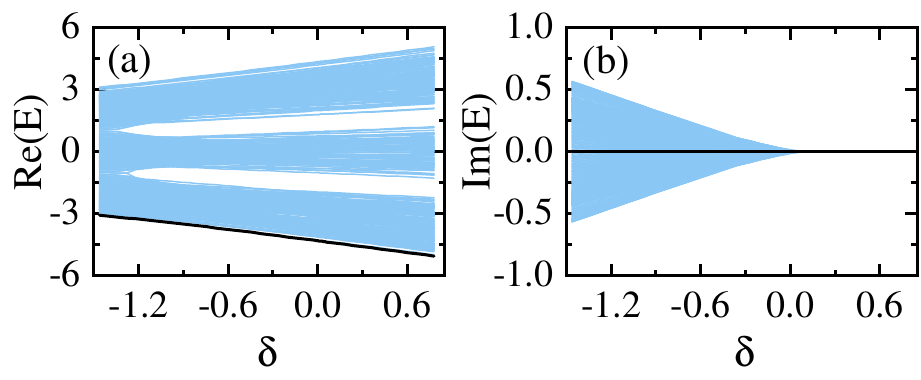}
  \vskip-3mm
  \caption{(a) Real and (b) imaginary parts of energy spectra of the model Eq. (1). The black curve that corresponds to the states with the lowest real part of the eigenenergy, where the energy spectrum is always real for all $\delta'$s. Here, we choose $g=0.5$, $\phi=0.2$, $\Delta=0.8$, and $L=377$ in the calculation.
  }
  \label{spectra}
\end{figure}

In Fig.~\ref{spectra}, we present the energy spectrum of the non-Hermitian DAA model.
Our findings indicate that the energy spectra predominantly exhibit real values when $\delta>0$, whereas they incorporate imaginary components when
$\delta<0$.
This suggests that the introduction of disorder disrupts the correspondence between the real-complex transition of the energy spectrum and the localization transition in the non-Hermitian AA model.
It can be observed that the characteristics of its energy spectrum are very similar to those of the non-Hermitian AA model~\cite{Zhai2022}, indicating that one-dimensional quasiperiodic systems are quite stable under perturbations by disorder~\cite{Zhang2019arxiv,zhangsx2018}.

\subsection{Character observables}
Here, we employ the $\xi$, IPR, and $\Delta E$ to explore the scaling law in the critical regions.

In the localized phase, the localization length $\xi$ for the non-Hermitian system is defined as~\cite{Sinha2019,Zhai2022}
\begin{equation}
\label{Eq:xiscaling}
   \xi = \sqrt{\sum_{n>n_c}^{L} [( n - n_c )^2 ] P_i},
\end{equation}
in which $P_i$ is the probability of the wavefunction at site $i$, and $n_c\equiv\sum nP_i$ is the localization center.
In the thermodynamic limit of $L\rightarrow\infty$, $\xi$ scales with the distance to the critical point $\varepsilon$ as
\begin{equation}
\label{Eq:xiscaling}
\xi\propto \varepsilon^{-\nu},
\end{equation}
where $\nu$ is a critical exponent.
For the pure non-Hermitian AA model, $\varepsilon=\delta$ and $\nu=\nu_\delta=1$ under both PBC and open boundary condition (OBC) ~\cite{Zhai2022,Zhaifphy}.

IPR is defined as~\cite{Bauer1990,Fyodorov1992}
\begin{equation}
\label{Eq:ipr}
{\rm IPR} = \frac{\sum_{j=1}^L||\Psi(j)\rangle|^4}{\sum_{j=1}^L||\Psi(j)\rangle|^2},
\end{equation}
where $|\Psi(j)\rangle$ is the right eigenvector.
For the extended phase, ${\rm IPR}$ scales as ${\rm IPR}\propto L^{-1}$.
For the localized state, ${\rm IPR}$ scales as ${\rm IPR}\propto L^0$.
At the critical point, $\rm IPR$ scales as
\begin{equation}
\label{Eq:IPRscalL}
{\rm IPR}\propto L^{-s/\nu},
\end{equation}
where $s$ is a critical exponent.
When $L\rightarrow\infty$, ${\rm IPR}$ scales with the distance to the critical point $\varepsilon$ as
\begin{equation}
\label{Eq:iprscaling2}
{\rm IPR}\propto \varepsilon^{s}.
\end{equation}
For the non-Hermitian AA model, $\varepsilon=\delta$ and $s=s_\delta=0.1197$~\cite{Zhai2022,zhai2021}.

At the critical point of localization transition, energy gap $\Delta E$ scales with the lattice size $L$ as
\begin{equation}
\label{Eq:GapscalL}
\Delta E\propto L^{-z},
\end{equation}
where $z$ is a critical exponent. For the non-Hermitian AA model, $\varepsilon=\delta$ and $z=z_\delta=2$~\cite{Zhai2022}.
When $L\rightarrow\infty$, energy gap $\Delta E$ scales as
\begin{equation}
\label{Eq:Gapscaling}
\Delta E\propto \varepsilon^{\nu z}.
\end{equation}

By taking into account the finite-size effect, the general scaling ansatz of a quantity $Y$ reads
\begin{eqnarray}
\label{Eq:generalscaling}
  Y(\varepsilon) &=& L^{y/\nu}f(\varepsilon L^{1/\nu}),
\end{eqnarray}
where $y$ is the critical exponent of $Y$ defined according to $Y\propto \varepsilon^{-y}$ when $L\rightarrow\infty$, and $f(.)$ is the scaling function.

\section{\label{criticalexponent}The critical exponents}
In this section, we study the scaling behaviors of the characteristic observables, and obtain the corresponding critical exponents for the pure non-Hermitian Anderson model and the non-Hermitian DAA model.

\subsection{The critical exponents for non-Hermitian Anderson model}

For the non-Hermitian Anderson model with $\delta=-2J_R$ in Eq.~(\ref{Eq:model}), the finite-size scaling functions for $\xi$, $\rm IPR$, and $\Delta E$ can be derived from Eqs.~(\ref{Eq:xiscaling}), (\ref{Eq:iprscaling2}), (\ref{Eq:Gapscaling}) and (\ref{Eq:generalscaling}).
The scaling function for $\xi$ reads
\begin{eqnarray}
\label{Eq:xiscalingNonDisorder}
  \xi &=& Lf_{1}(\Delta L^{1/\nu_A}),
\end{eqnarray}
where $\nu_A$ is the critical exponent for the non-Hermitian Anderson model, and $f_i$ is the scaling function.
Similarly, the scaling of $\rm IPR$ should satisfy
\begin{eqnarray}
\label{Eq:IPRscalingNonDisorder}
  {\rm IPR} &=& L^{-s_{A}/\nu_A}f_{2}(\Delta L^{1/\nu_A}),
\end{eqnarray}
where $s_A$ is the critical exponent of $\rm IPR$ for the non-Hermitian Anderson model.
The scaling function for $\Delta E$ reads
\begin{eqnarray}
\label{Eq:GapscalingNonDisorder}
  {\Delta E} &=& L^{-z_{A}}f_{3}(\Delta L^{1/\nu_A}),
\end{eqnarray}
where $z_A$ is the critical exponent of $\Delta E$ for the non-Hermitian Anderson model.

\begin{figure}[tbp]
\centering
  \includegraphics[width=1\linewidth,clip]{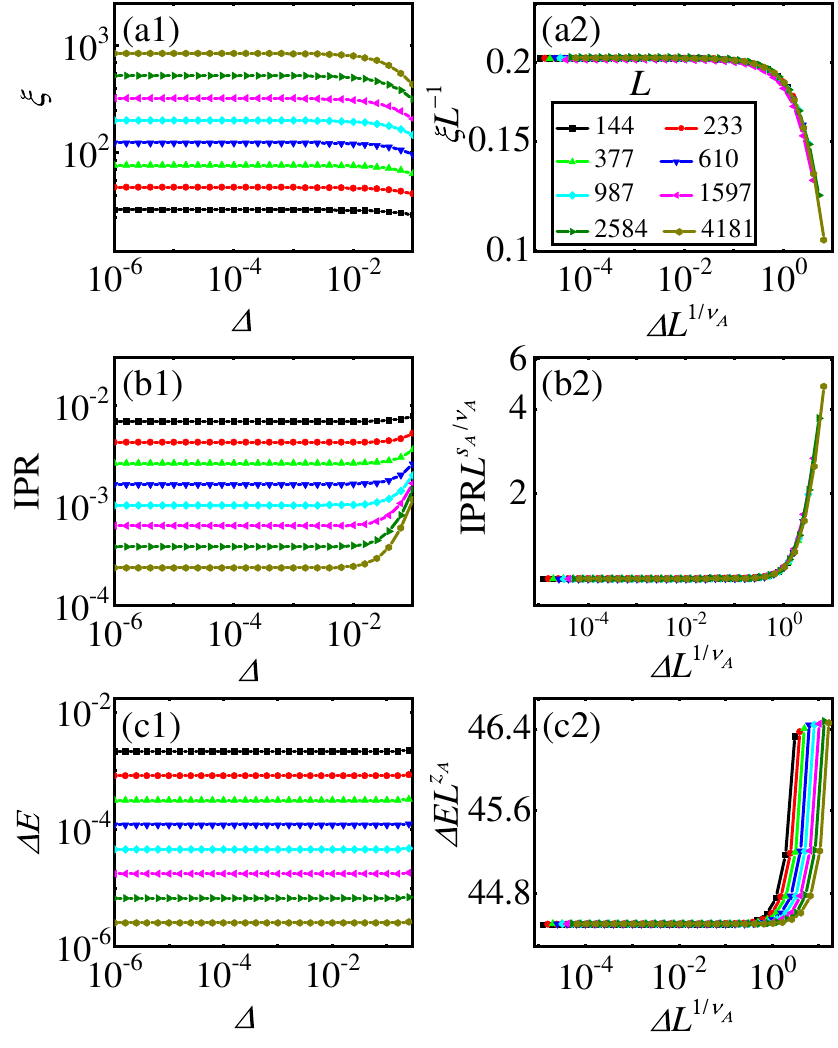}
  \vskip-3mm
  \caption{Scaling properties in the state with the lowest real part of the eigenenergy for the pure non-Hermitian Anderson model. The curves of $\xi$ versus $\Delta$ before (a1) and after (a2) rescaled for different $L'$s. The curves of IPR versus $\Delta$ before (b1)and after (b2) rescaled for different $L'$s. The curves of $\Delta E$ versus $\Delta$ before (c1)and after (c2) rescaled for different $L'$s. We use $g=0.5$, and the result is averaged for $1000$ samples of disorder. The double-logarithmic scales are used.}
  \label{Fig:puredisorder}
\end{figure}

To determine the critical exponents $\nu_A$, $s_A$ and $z_A$, we numerically calculate these three characteristic observables versus disorder strength $\Delta$, and rescale them according to Eqs.~(\ref{Eq:xiscalingNonDisorder})-(\ref{Eq:GapscalingNonDisorder}), respectively, based on some trial values for these variables.
In Fig.~\ref{Fig:puredisorder}(a1), we plot $\xi$ versus $\Delta$ for different $L$.
After rescaling $\xi$ and $\Delta$ as $\xi L^{-1}$ and $\Delta L^{1/\nu_A}$ with $\nu_A=1.99(9)$, we find that the rescaled curves collapse onto each other very well, as shown in Fig.~\ref{Fig:puredisorder}(a2).
The error estimation method employed is identical to that described in Ref. \cite{Xuan2022}, which relies on the observation of a deviation between the rescaled curves occurring when $\nu_A$ lies outside the range of 1.90 to 2.08.
In Figs.~\ref{Fig:puredisorder}(b1) and (b2), the numerical results of $\rm IPR$ versus $\Delta$ before and after rescaling to Eq.~(\ref{Eq:IPRscalingNonDisorder}) are plotted.
We find that collapse of the rescaled curves is best when $s_A=1.99(1)$, by setting $\nu_A=1.99(9)$ as an input.
The numerical results of $\Delta E$ versus $\Delta$ and the rescaled curves of $\Delta EL^{z_{A}}$ versus $\Delta L^{1/\nu_A}$ are plotted in Figs.~\ref{Fig:puredisorder}(c1) and (c2).
By setting $\nu_A=1.99$(9) as an input, we find that the best collapse of these rescaled curves appears when $z_A=2.00(1)$.

Hence, the critical exponent set of the 1D non-Hermitian Anderson model is obtained as $(\nu,s,z)=(\nu_A,s_A,z_A)=(1.99,1.99,2.00)$.
It is noteworthy that the  values of these critical exponents deviate from those observed in the 1D Hermitian Anderson model.~\cite{Wei2019,Xuan2022}.
This discrepancy indicates that non-Hermitian and Hermitian Anderson models are categorized under distinct universality classes.

Accurately estimating the critical exponents as well as the precision of those estimations is crucial for understanding the critical properties of the Anderson transition. Slevin and Ohtsuki proposed a correction method for finite-size scaling at the Anderson transition, which provides another possibility to further explore the reliability of numerical analysis results~\cite{Slevin1999}.

\subsection{The critical exponents for the non-Hermitian DAA model}

For the non-Hermitian DAA model, the critical exponents along the $\delta$ and $\Delta$ directions should be different, since $\delta$ and $\Delta$ are two distinct relevant directions.
The non-Hermitian DAA model returns to the non-Hermitian AA model when $\Delta=0$, hence, the critical exponents along the direction of $\delta$ should be $(\nu,s,z)=(\nu_\delta,s_\delta,z_\delta)=(1,0.1197,2)$~\cite{Zhai2022}.
The scaling exponents along the $\Delta$ direction $(\nu,s,z)=(\nu_\Delta,s_\Delta,z_\Delta)$ should be a distinct set from those in the non-Hermitian Anderson model.
This is due to the nature of the critical point at $(\Delta,\delta)=(0,0)$ being different from the extended state scenario.

\begin{figure}[tbp]
\centering
  \includegraphics[width=0.76\linewidth,clip]{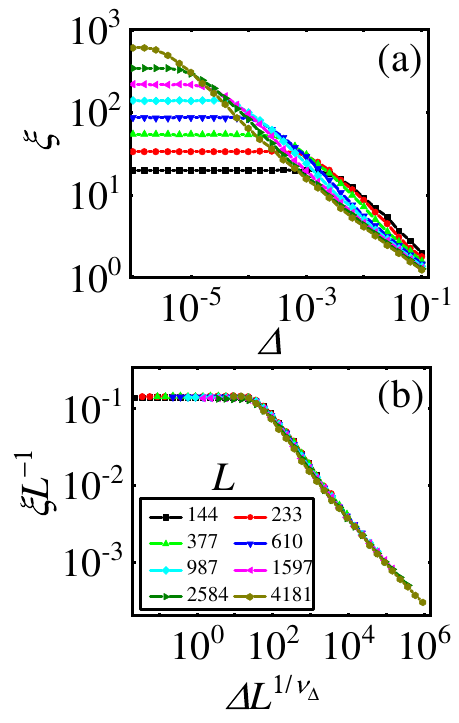}
  \vskip-3mm
  \caption{(a)Curves of $\xi$ versus $\Delta$ for the non-Hermitian DAA model at $\delta=0$ for various $L'$s.
  (b)The rescaled curves of $\xi L^{-1}$ versus $\Delta L^{1/\nu_\Delta}$ according to Eq.~(\ref{Eq:xiscalingNHDAA1}).
  We use $g=0.5$, and the result is averaged for $1000$ samples.
  The double-logarithmic scales are used.}
  \label{estimate_NuD}
\end{figure}

Along the $\Delta$ direction, localization length $\xi$ scales with $\Delta$ as $\xi\propto \Delta^{-\nu_\Delta}$.
Similarly, by taking into account the finite-size effect, the scaling ansatz of $\xi$ reads
\begin{equation}
\label{Eq:xiscalingNHDAA1}
\xi=Lf_4(\Delta L^{1/\nu_\Delta}).
\end{equation}
The curves of $\xi$ versus $\Delta$ for various $L$'s at $\delta=0$ are plotted in Fig.~\ref{estimate_NuD}(a).
After rescaling $\xi$ and $\Delta$ according to Eq.~(\ref{Eq:xiscalingNHDAA1}), we find that the rescaled curves exhibit a quite good collapse onto each other when $\nu_\Delta=0.52$(2), as plotted in Fig.~\ref{estimate_NuD}(b).
This new critical exponent $\nu_\Delta$ indicates that the disorder contributes a new relevant direction at the non-Hermitian AA critical point.

\begin{figure}[tbp]
\centering
  \includegraphics[width=0.8\linewidth,clip]{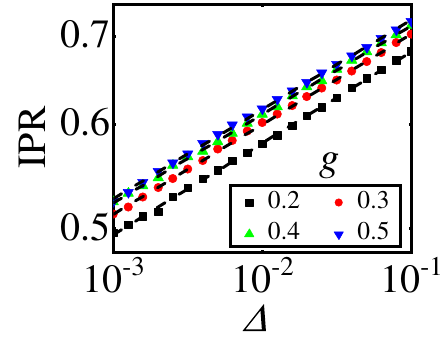}
  \vskip-3mm
  \caption{Curves of $\rm{IPR}$ versus $\Delta$ for the non-Hermitian DAA model at $\delta=0$ for various $g'$s.
  We use $L=4181$ and the result is averaged for $1000$ samples. The dashed lines are fitting lines. The fitting error of the exponent is $\pm 0.0004$.
   The double-logarithmic scales are used.}
  \label{IPR_fit}
\end{figure}

From Eq.~(\ref{Eq:GapscalL}), it is shown that $\Delta E\propto L^{-z_\delta}$ and $\Delta E\propto L^{-z_\Delta}$ should be applicable simultaneously. Hence, we have $z_\Delta=z_\delta=2$.
While for $\rm IPR$, the simultaneous applicability of Eq.~(\ref{Eq:IPRscalL}) in both directions provides the following relationship
\begin{eqnarray}
   s_\Delta=s_\delta \frac{\nu_\Delta}{\nu_\delta}.
\end{eqnarray}
Therefore, along the $\Delta$ direction, $\varepsilon=\Delta$ and $s=s_{\Delta}$, Eq.~({\ref{Eq:iprscaling2}}) becomes
\begin{eqnarray}
  {\rm IPR}\propto\Delta^{s_\Delta}= \Delta^{s_\delta\nu_\Delta/\nu_\delta}.
\end{eqnarray}
By studying the scaling properties of $\rm IPR$ with $L\rightarrow \infty$, we can determine $s_{\Delta}$ and further verify the critical exponent $\nu_\Delta$.

In Fig.~\ref{IPR_fit}, $\rm IPR$ versus $\Delta$ at $\delta=0$ for different $g$ values are plotted.
The lattice size is $L=4181$, which is large enough so that the size effects are tiny.
We find that the plots of $\rm IPR$ versus $\Delta$ are parallel lines in the double-logarithmic coordinates.
Notably, the average slope of these lines is $s_{\Delta}=0.0642$ with the fitting error of the exponent being $\pm 0.0004$, aligning close to the theoretical prediction value of $s_{\Delta}=s_\delta\nu_\Delta/\nu_\delta=0.0623$(24) by setting $\nu_\Delta=0.52$(2) as an input.
This consistency also confirms the correctness of the value of $\nu_{\Delta}$ within the error bar.
In addition, since we have calculated the results for different $g$ values and they all fit well with the theoretical predictions, this indicates that the obtained exponent is universally applicable to the non-Hermitian DAA model.

Thus, we obtain the set of critical exponents for the non-Hermitian DAA model in the $\Delta$ direction as $(\nu,s,z)=(\nu_\Delta,s_\Delta,z_\Delta)=(0.52,0.0642,2)$.
These exponents are different from that of the Hermitian DAA model~\cite{Xuan2022}.
This also indicates that the non-Hermitian DAA and the Hermitian DAA models belong to different universality classes.

\begin{figure}[h]
\centering
  \includegraphics[width=1\linewidth,clip]{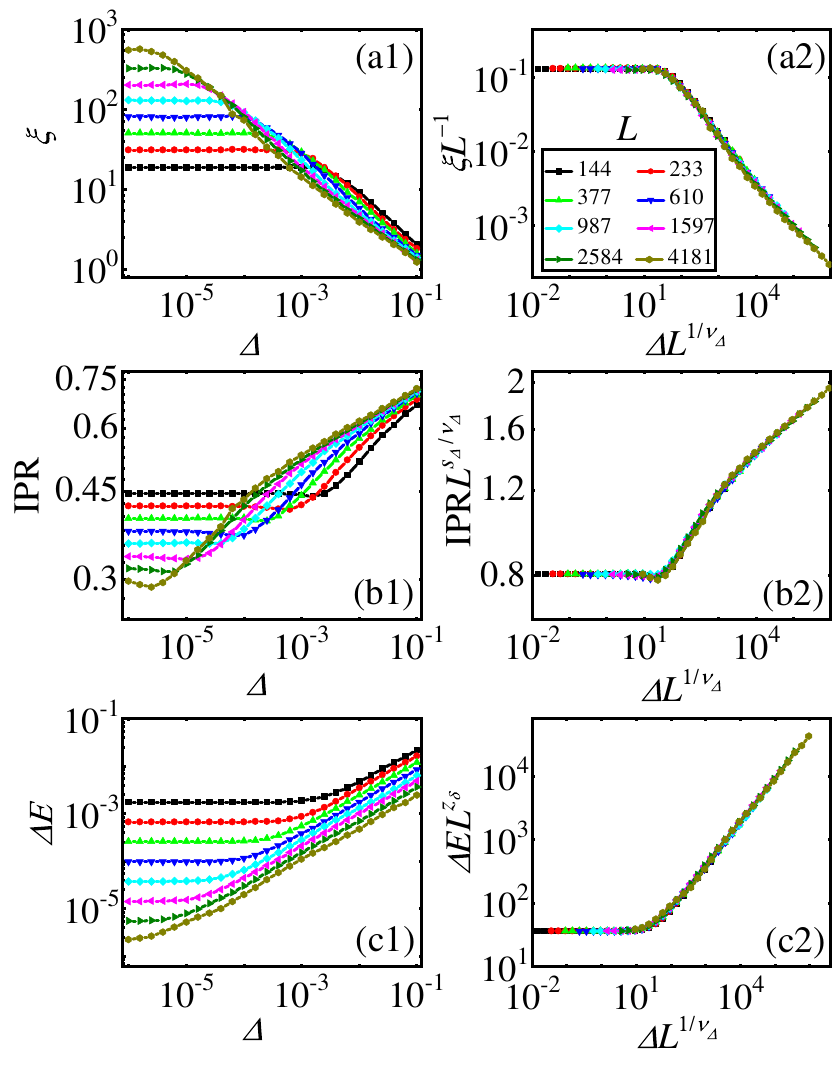}
  \vskip-3mm
  \caption{Scaling properties in the state with the lowest real part of the eigenenergy for fixed $\delta L^{1/\nu_\delta}=1$.
  The curves of $\xi$ versus $\Delta$ before (a1) and after (a2) rescaled for different $L'$s.
  The curves of $\rm IPR$ versus $\Delta$ before (b1) and after (b2) rescaled for different $L'$s.
  The curves of $\Delta E$ versus $\Delta$ before (c1) and after (c2) rescaled for different $L'$s.
  Here, $g=0.5$, and the result is averaged for $1000$ samples. The double-logarithmic scales are used.}
  \label{dL1}
\end{figure}

\begin{figure}[htbp]
\centering
  \includegraphics[width=1\linewidth,clip]{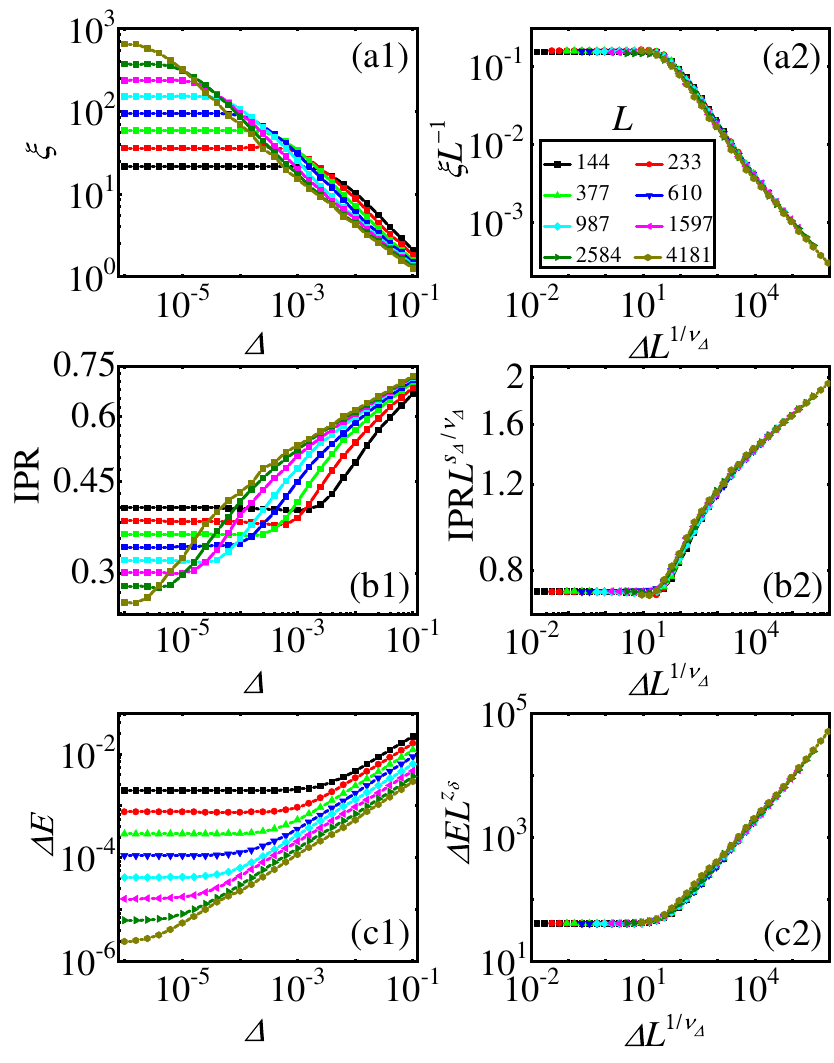}
  \vskip-3mm
  \caption{Scaling properties in the state with the lowest real part of the eigenenergy for fixed $\delta L^{1/\nu_\delta}=-1$.
  The curves of $\xi$ versus $\Delta$ before (a1) and after (a2) rescaled for different $L'$s.
  The curves of $\rm IPR$ versus $\Delta$ before (b1) and after (b2) rescaled for different $L'$s.
  The curves of $\Delta E$ versus $\Delta$ before (c1) and after (c2) rescaled for different $L'$s.
   Here, $g=0.5$, and the result is averaged for $1000$ samples. The double-logarithmic scales are used.}
  \label{dLn1}
\end{figure}

\section{\label{Hybridscaling}Hybrid scaling properties around the critical point of the non-Hermitian DAA model}
In this section, we study the scaling properties around the critical point of the non-Hermitian DAA model.
In particular, a hybrid scaling law is proposed to characterize the scaling properties in the overlapping critical region constructed by the non-Hermitian DAA critical region and the Anderson localization transition.

\subsection{Full scaling forms in the critical regions}
By taking $\Delta$ and $\delta$ as scaling variables, the general finite size scaling forms of these three observables are
\begin{eqnarray}
\label{Eq:xifull}
  \xi &=& Lf_{5}(\delta L^{1/\nu_\delta},\Delta L^{1/\nu_\Delta}),\\
\label{Eq:IPRfull}
  {\rm IPR} &=& L^{-s_{\delta}/\nu_{\delta}}f_{6}(\delta L^{1/\nu_\delta},\Delta L^{1/\nu_\Delta}),\\
\label{Eq:DEfull}
  {\Delta E} &=& L^{-z_{\delta}}f_{7}(\delta L^{1/\nu_\delta},\Delta L^{1/\nu_\Delta}).
\end{eqnarray}
In the critical region of the non-Hermitian DAA model, the scaling functions Eqs.~(\ref{Eq:xifull}) to (\ref{Eq:DEfull}) should be applicable.

In the overlapping critical region where $\delta<0$, both the scaling functions of the non-Hermitian Anderson transitions and the non-Hermitian DAA model should play significant roles.
Here, to study the scaling behavior in this overlapping critical region, the following hybrid scaling law is proposed.
In a typical scenario in which the overlapping critical region is postulated to be composed of critical region A and critical region B, the hybrid scaling law proposes the following hypotheses:
First, within the overlapping critical region, the critical properties should be concurrently describable by the critical theories pertaining to both region A and region B.
Second, a constraint should be imposed between the scaling functions of both region A and region B.

It's worth noting that overlapping critical regions are a common phenomenon in condensed matter physics~\cite{Xuan2022,Liang2024,Yin2017,zhai2018,Hesselmann155157,Stephanov3746}, and this hybrid scaling law has a general and universal significance.
For instance, both the Hermitian DAA and AA-Stark models exhibit overlapping critical regions of localization transition, and they have also confirmed the correctness of this hybrid scaling law~\cite{Xuan2022,Liang2024}.
Similarly, in the study of the nonequilibrium dynamics in the Yang-Lee edge singularity, a hybrid Kibble-Zurek scaling has been proposed to describe the behavior of driven dynamics in overlapping critical regions~\cite{zhai2018,Yin2017}.

Here, we take the critical properties of $\xi$ to illustrate this hybrid scaling law.
According to this hybrid scaling law, both the scaling functions of $\xi$, i.e., Eqs.~(\ref{Eq:xiscalingNonDisorder}) and (\ref{Eq:xifull}), are applicable in the critical region where $\delta<0$.
Combining Eqs.~(\ref{Eq:xiscalingNonDisorder}) and (\ref{Eq:xifull}), the constraint between these two scaling functions should satisfy
\begin{eqnarray}
\label{Eq:constraint}
  f_5(\delta L^{1/\nu_\delta},\Delta L^{1/\nu_\Delta}) &=& f_1[{\Delta L^{1/\nu_\Delta}(\delta L^{1/\nu_\delta})^\kappa}],
\end{eqnarray}
where $\kappa\equiv\nu_\delta(1/\nu_A-1/\nu_\Delta)$.
We find that $\kappa$ includes both the critical exponents of non-Hermitian DAA model and non-Hermitian Anderson model, which gives the constraint between these scaling functions.

\begin{figure}[htbp]
\centering
  \includegraphics[width=0.7\linewidth,clip]{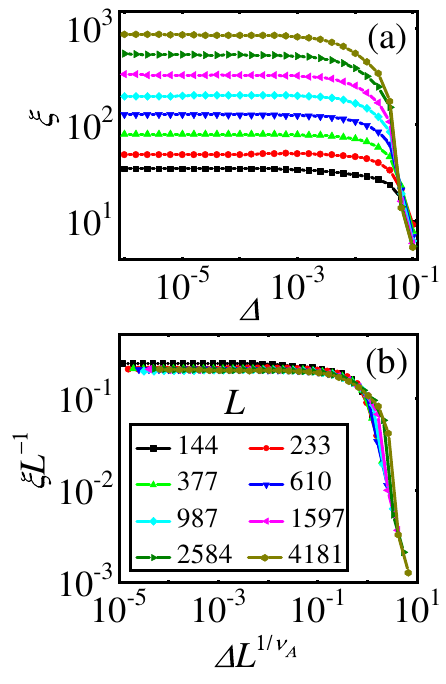}
  \vskip-3mm
  \caption{(a) The curves of $\xi$ versus $\Delta$ for different $L'$s at $\delta=-0.5$. (b) Rescaled curves of $\xi L^{-1}$ versus $\Delta L^{1/\nu_A}$ collapse onto each other. Here, $g=0.5$, and the result is averaged for $1000$ samples of $\phi$. The double-logarithmic scales are used.}
  \label{Disoverlapp}
\end{figure}

\begin{figure}[htbp]
\centering
  \includegraphics[width=0.7\linewidth,clip]{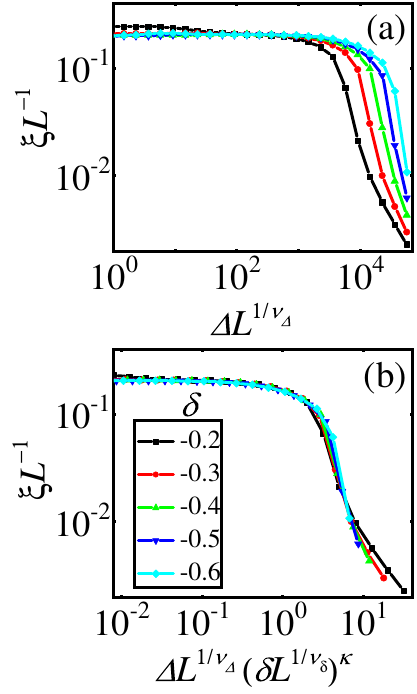}
  \vskip-3mm
  \caption{(a)The curves of $\xi L^{-1}$ versus $\Delta L^{1/\nu_\Delta}$ for different $\delta'$s at $L=987$. (b) Rescaled curves of $\xi L^{-1}$ versus ${\Delta L^{1/\nu_\Delta}(\delta L^{1/\nu_\delta})^\kappa}$ collapse onto each other. Here, $g=0.5$, and the result is averaged for $1000$ samples. The double-logarithmic scales are used.}
  \label{constrain}
\end{figure}

\subsection{Numerical results}

In this section, we numerically verify these scaling theories.
By fixing $\delta L^{1/\nu_\delta}$ at a constant value, Eqs.~(\ref{Eq:xifull})-(\ref{Eq:DEfull}) are first verified.
In Fig.~\ref{dL1}, the scaling properties of $\xi$, $\rm IPR$, and $\Delta E$ versus $\Delta$ for $\delta L^{1/\nu_\delta}=1$ are plotted.
After rescaling according to Eqs.~(\ref{Eq:xifull})-(\ref{Eq:DEfull}), the rescaled curves collapse onto each other well, confirming Eqs.~(\ref{Eq:xifull})-(\ref{Eq:DEfull}).
In the overlapping critical region with $\delta<0$, the similar numerical results for fixing $\delta L^{1/\nu_\delta}=-1$ are plotted in Fig.~\ref{dLn1}.
The collapse of the rescaled curves shown in Figs.~\ref{dLn1}(a2), (b2), and (c2) also confirms Eqs.~(\ref{Eq:xifull})-(\ref{Eq:DEfull}).

Then, we take $\delta=-0.5$ as an example to examine the applicability of Eq.~(\ref{Eq:xiscalingNonDisorder}) in the overlapping region, and the numerical results are plotted in Fig.~\ref{Disoverlapp}.
We find that the rescaled curves collapse onto each other well, indicating that Eq.~(\ref{Eq:xiscalingNonDisorder}) is still applicable in this overlapping region.
Therefore, numerical results in Figs.~\ref{dLn1}(a1) and (a2) as well as Fig.~\ref{Disoverlapp} confirm the first hypothesis of the hybrid scaling law.

The numerical results of $f_5=\xi L^{-1}$ as a function of $\Delta L^{1/\nu_\Delta}$ for various $\delta<0$ are plotted in Fig.~\ref{constrain}(a).
By rescaling $\Delta L^{1/\nu_\Delta}$ as ${\Delta L^{1/\nu_\Delta}(\delta L^{1/\nu_\delta})^\kappa}$, we find that the rescaled curves collapse very well, verifying Eq.~(\ref{Eq:constraint}) and the second hypothesis of the hybrid scaling law.

\section{\label{sum}summary}
In summary, we have studied the critical behaviors of the non-Hermitian DAA model, where non-Hermiticity is induced by nonreciprocal hopping.
The scaling functions of these quantities for the non-Hermitian Anderson model and the non-Hermitian DAA model have been obtained, and the critical exponents of these models have been determined.
We have discovered that the critical exponents of the non-Hermitian DAA differ from the non-Hermitian Anderson model and the non-Hermitian AA model, which indicates that disorder introduces a new relevant direction at the non-Hermitian AA critical point.
Critical properties in the critical region spanned by the disorder and quasiperiodic potentials have been explored in detail.
Especially, in the overlapping region constituted by the critical regions of the non-Hermitian DAA model and the non-Hermitian Anderson model, a hybrid scaling law is proposed and numerically verified.

On-site gain and loss is another important form of non-Hermiticity, and many studies have shown that the effects of this non-Hermiticity on the localization are quite different from those of nonreciprocal effects~\cite{Longhi2019,Xu2021}.
Therefore, as a potential extension of this paper, it is also worth investigating the non-Hermitian DAA model with on-site gain and loss.
Additionally, since the non-Hermitian disorder or quasiperiodic models are also highly sensitive to boundary conditions~\cite{Cai014201,Guo2021}, it is also necessary to discuss the critical properties of the non-Hermitian DAA model under OBC.
This could also be a possible extension of this paper.

\section*{Acknowledgments}
This work is supported by the National Natural Science Foundation of China (Grant No. 12274184), the Qing Lan Project, and  the Natural Science Foundation of the Jiangsu Higher Education Institutions of China (Grant No. 24KJB140008).

%
\end{document}